# Direct atomic fabrication and dopant positioning in Si using electron beams with active real time image-based feedback

S. Jesse,[1,2,*] B.M. Hudak,[1,3] E. Zarkadoula,[1,3] J. Song,[3] A. Maksov,[1,5] M. Fuentes-Cabrera,[2,4] P. Ganesh,[2,4] I. Kravchenko,[2] P.C. Snijders,[3] A.R. Lupini,[1,3] A. Borisevich,[1,3] S.V. Kalinin[1,2]

[1] The Institute for Functional Imaging of Materials, [2] The Center for Nanophase Materials Sciences, and [3] Materials Sciences and Technology Division, and [4] Computational Sciences and Engineering Division, Oak Ridge National Laboratory, Oak Ridge, TN 37831

[5] Bredesen Center for Interdisciplinary Research, University of Tennessee, Knoxville, Tennessee 37996, USA

Semiconductor fabrication is a mainstay of modern civilization, enabling the myriad applications and technologies that underpin everyday life. However, while sub-10 nanometer devices are already entering the mainstream, the end of the Moore's Law roadmap still lacks tools capable of bulk semiconductor fabrication on sub-nanometer and atomic levels, with probe-based manipulation being explored as the only known pathway. Here we demonstrate that the atomic-sized focused beam of a scanning transmission electron microscope can be used to manipulate semiconductors such as Si on the atomic level, inducing growth of crystalline Si from the amorphous phase, reentrant amorphization, milling, and dopant-front motion. These phenomena are visualized in real time with atomic resolution. We further implement active feedback control based on real-time image analytics to control the e-beam motion, enabling shape control and providing a pathway for atom-by-atom correction of fabricated structures in the near future. These observations open a new epoch for atom-by-atom manufacturing in bulk, the long-held dream of nanotechnology.

Research on transport phenomena in semiconductors in the late 40's at Bell Labs laid the foundation for many of the technologies that underpin modern civilization[1] and started the incessant drive for integration and miniaturization of electronic devices. Immediately after the demonstration of the solid-state transistor by Brattain, Bardeen, and Shockley[2], it was realized that the future lies in the integration of multiple devices, including transistors and memory elements, on a single crystal. While early strategies pursued the growth of compositionally-graded semiconductor crystals, it was the conceptual breakthrough by Noyes and Kirby that demonstrated the fabrication of in-plane structures in the form of the first integrated circuit, the accomplishment which landed them a Nobel Prize in 2015. Since then, the semiconductor industry has adopted a paradigm for fabrication based on a combination of 1D chemical steps (fabrication in the out of plane, or $z$-direction) such as oxidation, resist deposition, etching, etc. with 2D patterning steps (patterning in $xy$ plane) using light exposure. The combination of these steps in a predefined sequence, under well-defined conditions, has enabled the modern computer-based civilization, resulting in the present sub-10 nm semiconductor structure.

The undeniable success of present day semiconductor technology is belied by significant limitations. Device processing relies on mesoscopic transport and chemical reactivity, leading to rapid growth of stochastic phenomena and noise during fabrication. Shrinking device size combined with the discrete nature of atomic dopants leads to a large spread in device performance, which can be traced to different distinct (and uncontrollable) atomic configurations .[3] Applications ranging from micro- to nanomechanical systems necessitate the assembly of complex 3D structures, rather than densely integrated layers. These limitations are well-recognized in the semiconductor industry, and the emergence of techniques such as electron beam induced depositions and lithography or focused ion milling[4,5] is a direct response to these challenges. While lacking the parallel nature of photolithography, all of these techniques have developed into multibillion-dollar industries.

However, electron beam based techniques still lack the capability to fabricate materials down to the atomic level, and the need for such fabrication is by now well realized. In particular, quantum devices for large-scale implementation of quantum computing, single-spin magnetoelectronic devices, and scalable neuromorphic systems all require fabrication at the atomic level, including precise fabrication of crystalline layers down to single atomic planes, positioning of functional dopant atoms, and avoiding atomic-scale defects in the active region of the device and interconnects.[6,7] In other areas, the impact of these developments can be predicted. For example, in materials science and condensed matter physics the capability to create desired atomic configurations and explore their functional properties (e.g. via local electron spectroscopies) will yield a paradigmatic shift in our understanding of the underlying principles. In other areas, atomic level fabrication can provide pathways towards large scale fabrication of materials with predefined properties – e.g. by providing seed crystals of thermodynamically metastable phases that can be further grown in macroscopic crystals.

Despite this clear incentive, the current methodology for atomic fabrication today is the combined approach based on scanning tunneling microscopy manipulation and surface chemistry, harking back to experiments by D. Eigler and advanced by J. Lyding, M. Simmons,[6-8] and commercially by companies such as Zyvex[9] and NanoFactory[10]. In this approach, the ability of the scanning tunneling microscope to manipulate single atoms is combined with the precise control of surface chemistry (silicon passivation) to position dopants at preselected locations, interface with macroscopic electrodes, and stabilize with surface passivation layers that allow taking the fabricated structure outside of the ultra-high vacuum (UHV) chamber. However, the cost and throughput of this approach remains such that research grade, several qubit devices remain the only viable target application. Hence, the question remains – are there other strategies that can potentially enable atomic scale fabrication of semiconductors avoiding the throughput bottle neck of probe based fabrication?

Here, we demonstrate a novel method for semiconductor manufacturing: the use of the atomically focused beam of a scanning transmission electron microscope to control local material structure in the bulk with atomic precision. Through use of the electron beam, we can induce a broad variety of phenomena including amorphization, crystallization, and dopant atom motion that can be resolved *in-situ*, enabling real-time correction and editing of matter at the atomic level. As an illustration, we implemented a combination of e-beam control electronics and active machine vision based feedback to fabricate predefined crystalline Si patterns.

*Electron Matter Interactions in STEM*

Traditionally, STEMs have been perceived only as imaging or analysis tools, and any beam-induced modifications are considered undesirable beam damage. Yet in the last five years, it has been demonstrated that beam-induced modifications can produce appealing results, including formation and ordering of oxygen vacancies,[11] single defect formation and motion of extended defects in 2D materials,[12] beam-induced migration of single interstitials in diamond-like lattices,[13] and formation of single chemical bonds.[14] Remarkably, these changes often involve one atom or small groups of atoms, are potentially chemically selective, and can be monitored in real-time with atomic resolution,[15] opening pathways towards control.[16,17] This combination of atomic manipulation and (sequential) atomic-scale visualization was without precedent until we observed that the e-beam can induce the crystallization of certain amorphous materials including oxides such as $SrTiO_3$, a process we refer to as e-beam sculpting.[18] Notably, the interaction between the electron beam and amorphous matter was actively explored in the 1980's and 1990's, and e-beam crystallization of a number of important semiconductors such as $Si^{19-22}$ and $GaAs^{22-24}$ has been reported. However, these experiments lacked the capability to probe beyond mesoscopic level studies, and no attempts to actively direct and control the process were reported. Nonetheless, three key factors were established regarding beam-induced processes in semiconducting materials: there is a strong beam energy dependence controlled by the proximity to the knock-on threshold (roughly 145 keV for bulk Si), the interactions generally cannot be reduced to purely thermal processes, and under certain conditions these processes can result in both amorphization and crystallization of material. At that time, this approach did not appear to offer any significant advantages over classical semiconductor processing, and therefore was not extensively pursued.

Here, we use the atomically focused beam of a STEM operating at 200 kV to guide amorphous-crystalline transformation in Si at the atomic-plane level, including both forward and reverse transitions, and demonstrate beam-induced motion of dopant atoms that can be assembled in different configurations. As a model sample, we have chosen amorphous silicon grown on a crystalline Si substrate. The STEM image prior to e-beam crystallization is shown in Figure 1 (a). The boundary (marked with a dotted line) between crystalline and amorphous Si is clearly visible, and dopant atoms can be seen within the amorphous Si matrix.

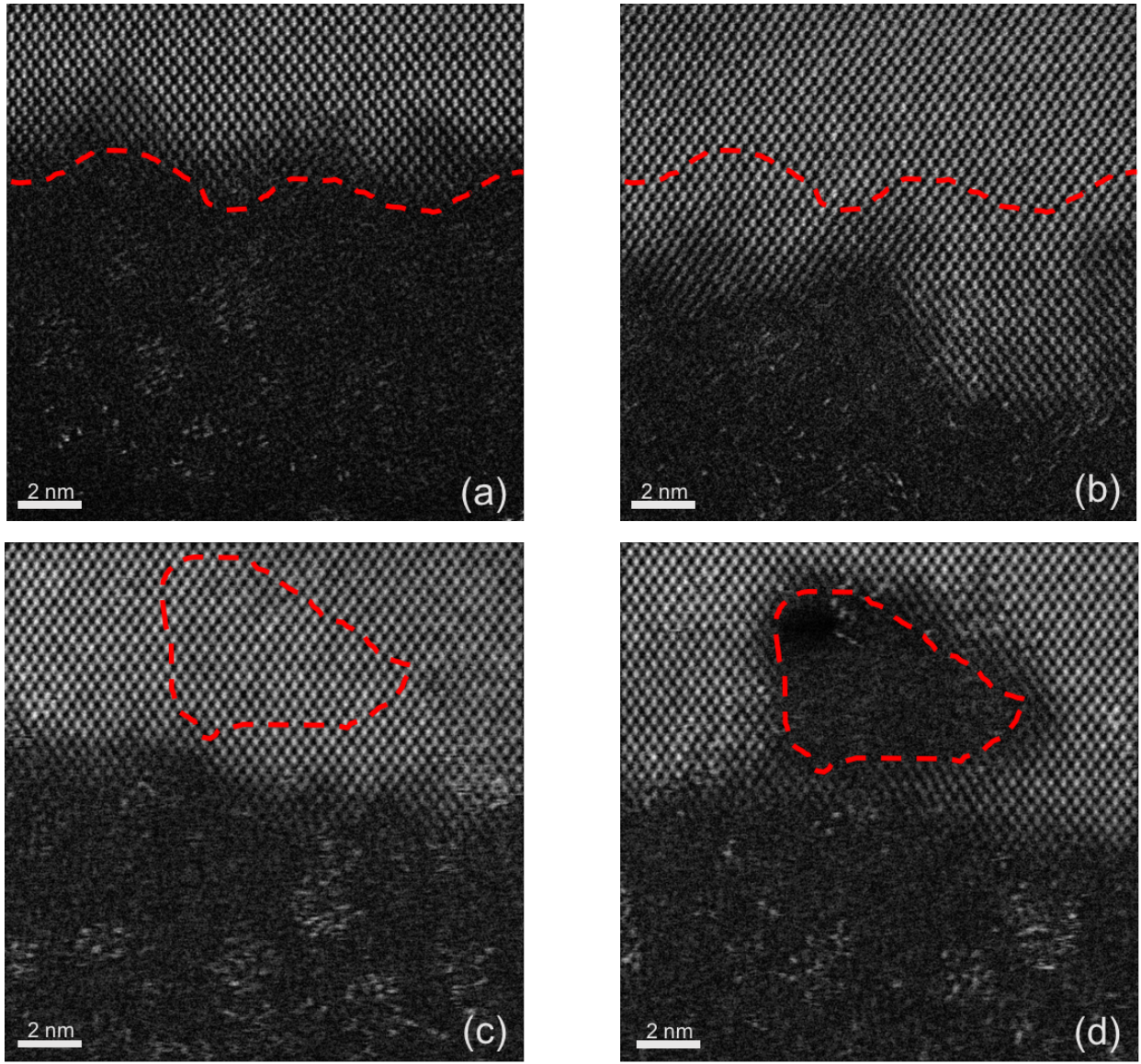

**Figure 1.** Beam induced transformations in Si using a 200 kV beam. (a) Before growth, (b) Crystallization of amorphous Si along the crystalline-amorphous interface (30 pA beam current, 200 kV). (c, d) Amorphization and subsequent drilling through of crystalline Si (139 pA beam current, 200 kV).

Figure 1 (b) shows the changes in the atomic structure after repeated scans of the image, with the slow scan direction perpendicular to the interface. There is a clear formation of crystalline Si extending into the amorphous region, in apparent epitaxial registry with the substrate. By increasing the beam current to 139 pA, compared to the nominal 30 pA conditions under which Figure 1(a-b) was obtained, we can drive the transformation from crystalline Si to amorphous Si, as shown in Figure 1(c-d). The increase in beam current results in the amorphization and subsequent drilling-through of crystalline Si.

We have further explored Si patterning over a range of beam current settings and scanning speeds, and demonstrate that it is possible to transition between crystallizing, amorphizing, and drilling/evaporating regimes by moving across these parameters. At 200 kV, the nominal current of the incident electron beam is 30-35 pA. At the nominal current setting and with a reduced scan speeds we are able to crystalize amorphous regions of Si. Increasing to an intermediate current of 75-80 pA and medium to high scan speed results in amorphization of crystalline Si. A high current mode of 140 pA and low to medium-high scan speeds leads to drilling or evaporating of the material. While the exploration of quantitative mechanisms behind the observed phenomena is a separate and complex topic requiring detailed studies[25,26], these observations clearly illustrate that both the fabrication and erasing materials regimes are open for experimental studies. Furthermore, the fact that beam-induced transitions between the phases can be reversible opens a tremendous field for further applications, from memory devices to reconfigurable electronics.

To obtain further insight into the structure of the newly formed crystalline Si, we perform comparative crystallographic image analysis[27,28]. In this method, a sliding window is scanned across the image, generating a stack of sub-images. The relevant 2D structure factors are calculated, and the resulting data set is linearly unmixed using non-negative matrix factorization (NMF). This procedure is ideally suited for differentiation of dissimilar crystalline phases, so we can apply it here to determine if the beam-crystallized Si grows with the same crystal structure as the crystal Si substrate. Unlike methods based on direct analysis of atomic positions, this method does not require high contrast images, i.e. unmixing is possible for cases where only lowest-order reciprocal lattice peaks are visible. We start by assuming that the initial image (Figure 2(a)) contains only two phases and perform unmixing for two endmembers. The resulting abundance maps, along with insets showing unmixed FFT endmembers, are shown in Figure 2(b, c). We then use non-negative least squares analysis (NNLS) in conjunction with discovered endmembers on the after-growth image (Figure 2(d)). Generated abundance maps clearly show that growth regions consist of the same crystallographic phase as the substrate (Figure 2(e, f)).

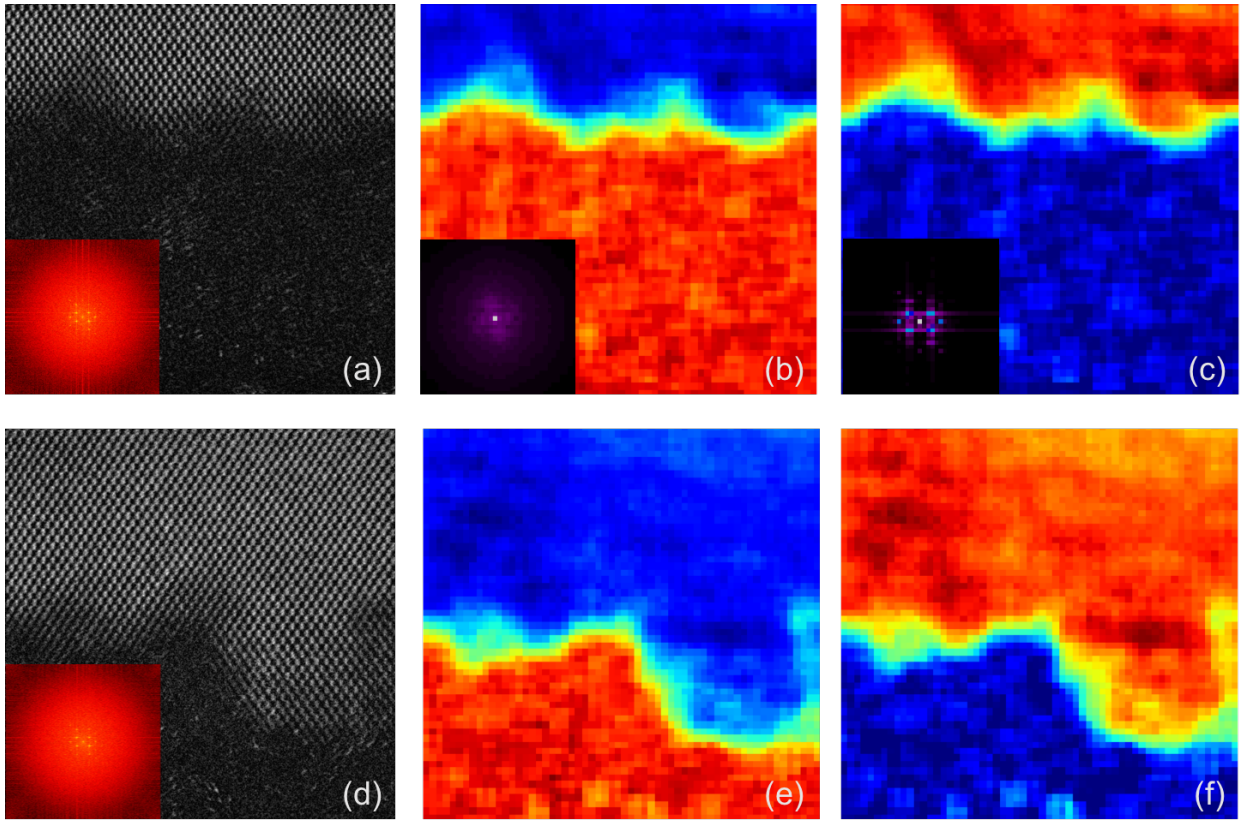

**Figure 2.** Atomic scale tracking of the local crystallinity. (a) Pre-growth image and its FFT. (b) Abundance map and endmember 1, recovered through NMF, corresponding to the amorphous region (c) Abundance map and endmember 2, recovered through NMF, corresponding to the crystalline region. (d) After-growth image and its FFT. (e) Abundance map for the amorphous region in after-growth image generated using NNLS. (f) Abundance map for the crystalline region in after-growth image generated using NNLS.

*Directed Crystallization*

      The combination of e-beam amorphization/crystallization and real-time observation of these transformations suggests the potential for real-time feedback. based on simultaneously monitoring the images to keep track of the beam-induced changes. The level of crystallization can be monitored and controlled to produce structures with a desired geometry. Here, a custom feedback and control system has been developed to guide the atomic layer by atomic layer movement of the crystal-amorphous (CA) interface (either as crystallization into the amorphous region or amorphization in the reverse direction). The system operates by scanning the electron

beam parallel to the CA interface and simultaneously capturing the bright field and/or dark field signals of the STEM during these linear scans. A single line scan can be used to determine the degree of local crystallinity by calculating the amplitude portion of the 1D fast Fourier transform (FFT) of the STEM image signal – a line scan across an amorphous region will result in a relatively featureless FFT, whereas a line scan across a crystalline region will yield easily identifiable peaks corresponding to the average spacing between atomic columns. Controlled movement of the CA interface is achieved by using the magnitude and location of these peaks as a feedback signal to move the electron beam appropriately. That is, if one intends to advance the crystalline domain into the amorphous region then: (1) repeated identical line scans across the CA interface are performed to both induce crystallization and assess the degree of local crystallinity, (2) when the degree of crystallinity reaches a pre-determined set-point (i.e. an atomic layer of atoms has transformed from amorphous to crystalline), the line scan is advanced approximately half a unit cell into the amorphous region, and the process is continued. This process is illustrated in Figure 3.

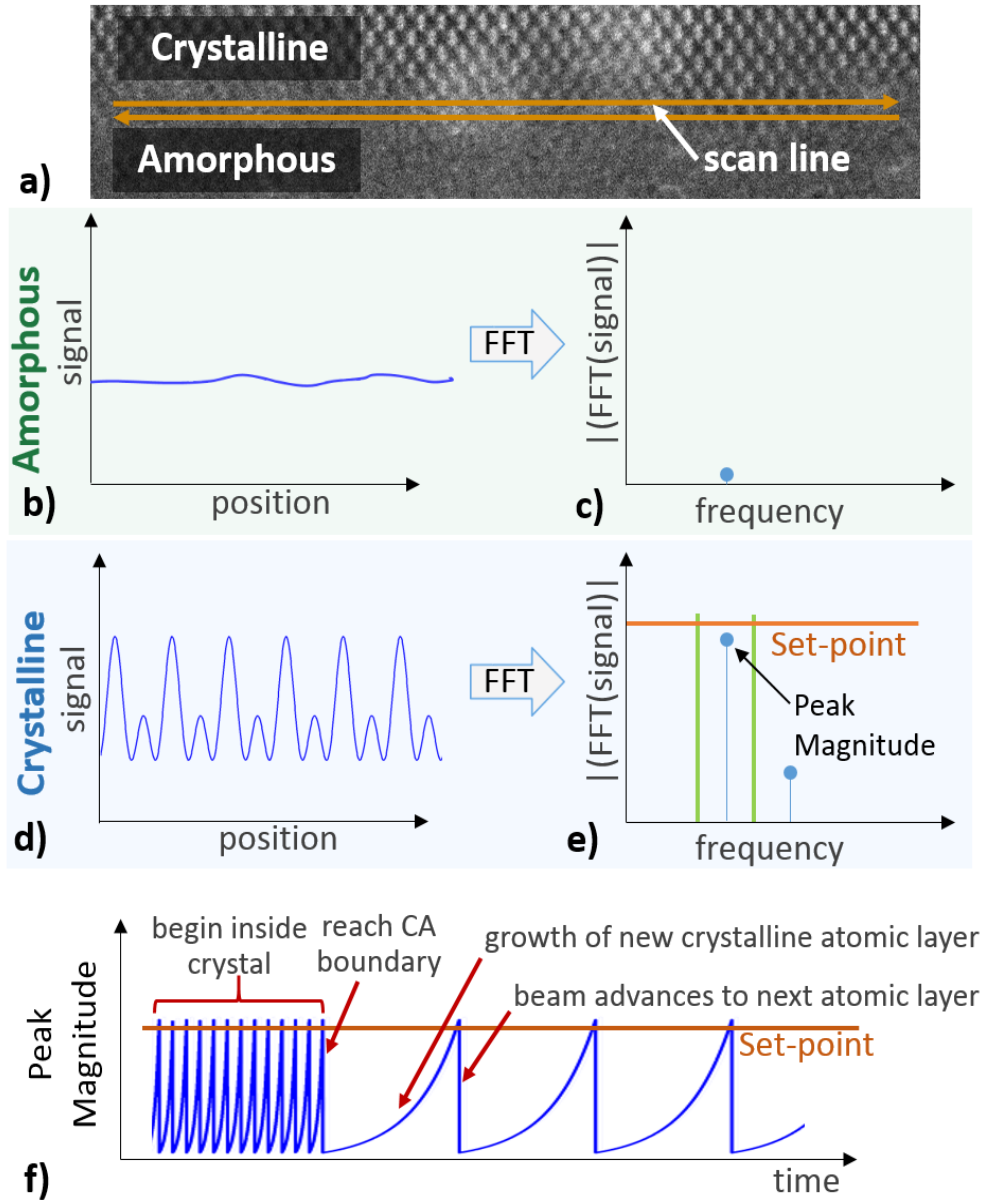

**Figure 3:** Beam directed crystallization process. a) STEM image of CA interface and location of scan path during crystallization. Schematic depiction of the patterning process. (b) Spatial domain and (c) Fourier domain plots of detector signal in amorphous region. (d) Spatial domain and (e) Fourier domain plots of detector signal in crystalline region. The amplitude of a specific Fourier peak is tracked relative to the set-point. When amplitude exceeds the set-point, the beam is advanced ½ a unit cell into the amorphous region. (f) Plot of the peak amplitude from (e) as a function of time during directed crystal growth from a starting point within the crystal and while the CA front proceeds forward.

A similar procedure can be used to advance the CA interface into the crystalline region for controlled amorphization as well. The differences being that beam conditions are selected to cause the crystalline portions to amorphize, and the feed-back condition becomes that advancement of the beam only occurs when peak magnitude *drops below* a specified set-point.

### *Direct Growth of Doped Silicon*

To explore the fundamental processes during beam induced crystallization and amorphization, the experiments were repeated for Si containing dopant atoms. For most modern semiconductors, dopants are essential to obtain the desired electrical characteristics; however, direct visualization of single dopants remains a technical challenge. In these samples, a layer of Bi atoms was deposited at the CA interface. Single Bi are highly visible in STEM Z-contrast (Figure 4). By scanning the beam in a similar fashion as described above to induce crystal growth, it is possible to induce motion of bismuth atoms perpendicular to the fast-scan direction of the beam. This is demonstrated under various conditions in Figure 4. Figures 4(a) and (b) show, respectively, the before and after images of crystal growth where the beam-induced growth process was initiated between the CA interface and the region of high concentration dopants, propagating in [110] crystallographic direction. Figure 4 (b) shows the resulting crystal growth and the apparent movement of a few dopant atoms along the crystal growth direction. Note here, that at this low concentration of dopants, crystal growth appears to be unimpeded and in fact seems to allow for larger-scale structures than in the undoped case described above (perhaps due to some strain relief). For comparison, in (c) the beam induced growth is initiated deeper within the crystal than the layer with high dopant concentration, and thus a large number of dopant atoms are displaced. In this case however, crystal growth stopped progressing after several nanometers, presumably due to poor crystallographic compatibility between Bi and Si.

Further investigations of growth and dopant motion was performed by progressing in the [111] direction. Similar behavior was exhibited in (e,f) as in (c,d), namely the crystal growth and dopant motion proceeded together and abruptly stopped when the local dopant concentration reached a critical value. However, with dopant atoms cleared out of the way in (f), it was possible to induce crystal growth perpendicular to the original [111] direction as shown in (g). Figure 4(h) and (i) show that the same scanning pattern can also be used to move dopants deeper into the crystal. Notably, the dopant front appears very sharp in both cases, and no fronts form on the sides of the growing crystal. This suggests that the hopping/relaxation time for Bi atoms after being activated ("knocked") by the electron beam is closer to the scan time of an entire line in the fast scan direction rather than a pixel in that line. A process at this timescale can be precisely controlled by the electron beam, directly illustrating the feasibility of controlled atom-by-atom motion. Interestingly, this does not appear to be the case when the slow scan direction is along the [110] crystallographic axis. There is no apparent sharp front, and the resulting dopant profile resembles a dome more than a line. This difference in behavior for different crystallographic axes suggests that [111] is the preferential hopping direction for Bi atoms and highlights the importance of understanding the underlying mechanism and development of a predictive modeling approach for achieving reproducible results.

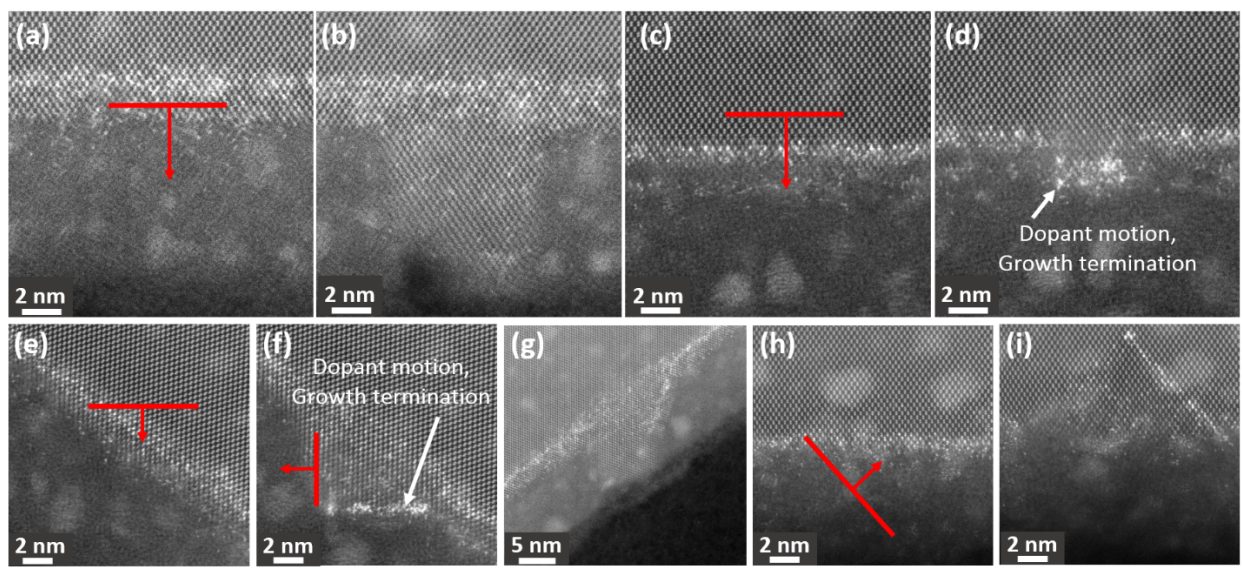

**Figure 4.** Controlled crystallization of amorphous Si and dopant movement. Red arrows illustrate the direction of the slow scan axis for growth, while the red lines at the base indicate scan width. Panels with red arrows are the "before" images of growth, and the following panels are the "after images; panel (f) serves as both for the (e-g) sequence. Note that the field of view is rotated and magnification is changed between panels (f) and (g) to provide better overview of the changes.

*Modelling Electron Beam Induced Transformations*

To gain further insight into the observed phenomena further probe the observed phenomena, we consider the effects of the electron beam on the solid. Generally, the energy transfer between a high-energy particle and a solid includes two primary components: losses to the electronic subsystem and direct interactions between high-energy particles and nuclei (knock-on). The knock-on interaction can result in damage when the kinetic energy that can be transferred in a single collision is larger than the energy barrier to displace an atom in the solid. Notably, in amorphous materials the binding energies are broadly distributed, allowing for a broad distribution of knock-on thresholds. For sufficiently high particle energy, multiple event cascades can be initiated. Similarly, in a material with finite thickness, when the knock-on interaction occurs a few layers away from a material surface, ejection of surface atoms is possible. Additionally, the energy barrier will usually be significantly lower for surface atoms, primarily because of the reduced number of bonds[25,26].

A powerful model for beam-induced changes in materials includes non-equilibrium heating, when the two subsystems – atomic nuclei and electronic – develop different temperatures, thus being in non-equilibrium conditions. Depending on the temperature difference between the two subsystems, energy that is transferred to the electrons can subsequently be transferred to the lattice atoms via the electron-phonon interactions until equilibrium is reached, where it diffuses further through the atoms. This mechanism is described by the two-temperature (2T) model[29-32]. In the 2T model, the evolution of the electronic and the atomic temperatures are described separately, using a set of heat diffusion equations, one for the electronic and one for the atomic system. The energy exchange between the two subsystems depends on the temperature difference between them, and the strength of this interaction is expressed with the electron-phonon coupling parameter $g$. For the case of silicon, a combination of results from irradiation experiments[36] combined with the inelastic thermal spike model[33,34], molecular dynamics and DFT computational[35,36] and model and numerical approaches[37-37] determine the values of the 2T model parameters. From this, $g$ is calculated to be $1.8 - 5 \times 10^2$ W/cm$^3$/K[33,34,37], using the known values for the lattice specific heat and conductivity[38], and the electronic specific heat and diffusivity[36].

Given the uncertainties in these parameters for amorphous solid, here we modeled the induced crystallization assuming that the electron beam creates a local temperature within a small volume of material. To explore this behavior, we developed a molecular dynamics model for amorphous Si in contact with crystalline silicon. The heated region (20 Å x 10 Å x 108 Å), representing the local volume heated by the beam, was initiated at the CA interface, and slowly moved into the amorphous region. Once the temperature inside the block reached 1300 K, crystallization began in regions close to the interface and moved upwards, terminating approximately at the [111] face, resulting in a pyramidal-like front. After 1 ns, the "beam" was moved 5 Å further into the amorphous region, and a block of the same size was again heated to 1300 K. The rest of the system, which now includes half of the previously crystallized block, was kept at 300K. This process was repeated several times until the crystal front reach about half the size of the amorphous sample (50 Å).

In order to differentiate between crystalline and amorphous phases we use the tetrahedral order parameter[39] to describe coordination state of each atom at 20 ps intervals according to the formula

$$q_t = 1 - \frac{3}{8}\sum_{i<j=1}^{4}(cos\theta_{ij} + \frac{1}{3})^2,$$

where $\theta_{ij}$ is the angle between an atom and its two nearest neighbors. The resulting parameter is in the range between zero, indicating an amorphous phase, and one, indicating a crystalline phase. However, within the amorphous phase we also observe multiple small momentarily crystalline regions. For each analyzed frame, we construct the matrix of tetrahedral order parameters for each atom and its corresponding 12 nearest neighbors. We use a k-means clustering algorithm[40] on the first frame to train the classifier, and use it to predict phases in the subsequent simulation frames. Figure 5 demonstrates the application of this algorithm to the simulation data, showing only atoms belonging to a crystal. This corroborates evidence from the experiment that the growth belongs to the same crystalline phase as the substrate, since we use pre-growth data as a training set, and the growth is classified by the unsupervised algorithm as the same crystal as the original substrate.

We observe that crystallization does occur mostly inside the heated region, forming characteristic pyramidal growth pattern in the beginning, and becoming slightly wider at the top of the growth region later (Figure 5 insets). Along the length of the crystallization front, we observe a characteristic wave-like pattern, which gets amplified as the heated region moves further away from the substrate.

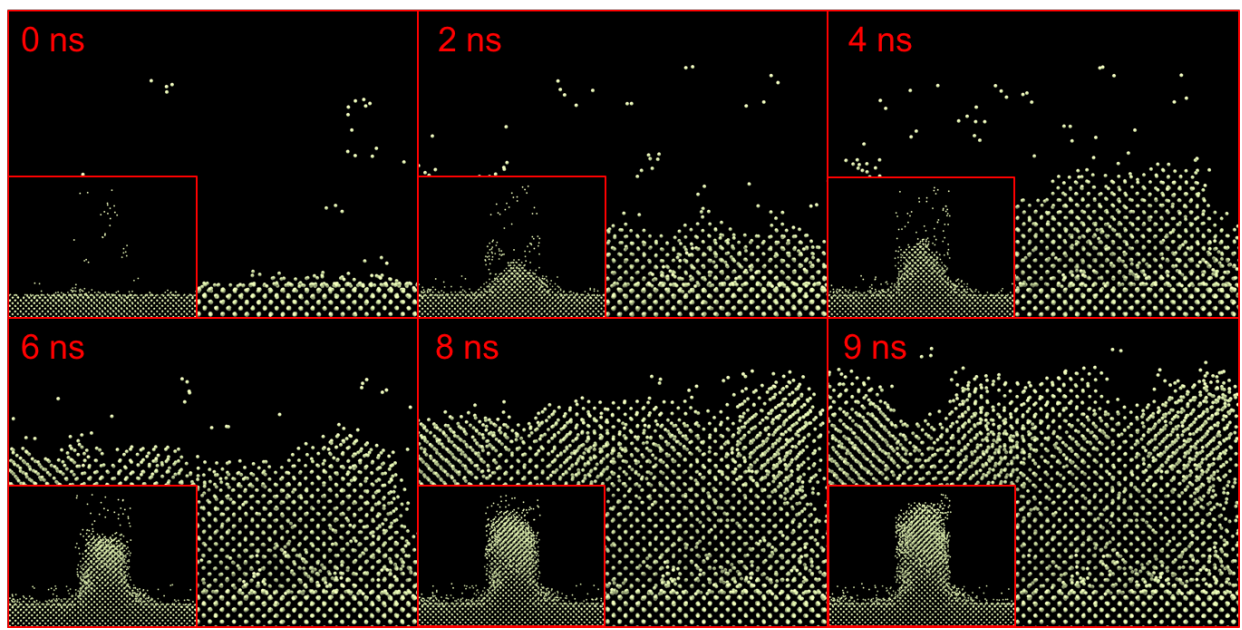

**Figure 5.** Timeline showing views of crystallization front in molecular dynamic simulation of Si crystallization.

The simulation, despite being simplified, reproduces much of the experimental behaviors, including the tendency to form triangles (pyramids) terminating at [111] planes. This model further offers evidence of roughening, with the brightness of the atomic columns decreasing quickly the further we get from the original boundary, observed both in experiment and theory. Further model development necessitates inclusion of realistic time effects, since presently the time scale is ~ns, as compared to experimental 10's of seconds. This behavior can be linked to a higher heating rate in MD compared to the experiment, and also allows to compensate for mismatch in timescales. Secondly, we aim to include the contribution of knock-on effects. However, even this simple model provides insight into the morphologies of the experimentally grown structures.

**Perspectives**

Since the early days of nanotechnology revolution, the development of realistic pathways for atom-by-atom fabrication was seen as the key and enabling step to bring its promise into reality. This requirement is most acutely felt now, with the industry pace given by Moore's Law getting to the single-digit nanometer device size, and with new devices based on behaviors of a single atom, such as for quantum computing, rising to the forefront of research and development.

The atomic manipulation of Si, the most important industrial semiconductor, demonstrated here, marks a key step in this direction. Remarkably, the capability of the electron beam to crystallize, amorphize, remove material, and controllably move dopant atoms fronts, even under the limitations of microscopes primarily designed to image materials rather manipulate matter, suggests that there may exist enormous potential to shape and direct matter on the atomic level.

While predicting all the opportunities enabled by the potential of STEM to manipulate and control matter at the atomic level will be complex, here we comment on the likely pathways for the development of the field in the next several years. First and foremost, the real-time feedback system implemented here can be expanded to include more complex forms of image analytics, e.g. switching between "modification" and imaging modes. Here, the use of compressed sensing[41] and related approaches could be instrumental in disambiguating low-dose non-invasive and high-dose modification regimes. Secondly, using a full 2D readout from a fast Ronchigram detector instead of simply a HAADF intensity reading can provide a feedback signal that can be used to determine when a desired transformation has occurred while the beams remains at a single location. Third, further development of precise control systems that are capable of high-speed and high-veracity beam positioning by compensating for beam scanning non-idealities (such as phase lag and frequency dependent gains) will be required.

These studies also call for extensive theoretical exploration of beam-induced effects in solids on the atomic level, at time scales spanning ultrafast electron transit times to the seconds of the induced structural relaxations. Evolution of electronic, lattice, and concentration fields and their interdependence need to be considered in detail. We believe that the results shown above represent an important step towards full experimental control and theoretical understanding of the process.

**Acknowledgments**

This research was conducted at and partially supported (SJ, IK, MFC, GP, SVK) at the Center for Nanophase Materials Sciences, which is a US DOE Office of Science User Facility. Part of this work was (B.M.H., J.S., P.C.S) supported by the Laboratory Directed Research and Development program of the Oak Ridge National Laboratory, managed by UT-Battelle, LLC, for the U.S. Department of Energy. A.B., A.R.L. and EZ were supported by the U.S. Department of Energy, Office of Science, Basic Energy Sciences, Division of Materials Science and Engineering. A. M. acknowledges fellowship support from the UT/ORNL Bredesen Center for Interdisciplinary Research and Graduate Education. This research used resources of the Oak Ridge Leadership Computing Facility at the Oak Ridge National Laboratory, which is supported by the Office of Science of the U.S. Department of Energy under Contract No. DE-AC05-00OR22725.

**Materials and methods**

**STEM experiment:**

A Nion UltraSTEM aberration-corrected scanning transmission electron microscope, operating at 200 kV was used in this work. The nominal convergence angle was 30 mrad. High angle annular dark field (HAADF) images were acquired on a detector with an inner angle of 63 mrad. The nominal probe current is 30 pA, and the current was varied during experiments to direct crystallization and amorphization.

**Modelling:**

A crystalline Si sample with 64000 atoms was relaxed at 0 K using the Stillinger-Weber potential given in Sastry et al.[42] Heating up this system to a temperature of 2000 K at constant volume using the canonical, NVT, ensemble and the Nosé-Hoover thermostat, created an amorphous sample with 64000 atoms. (NVT was chosen here so as to ensure that both the amorphous and the crystal sample upon which the former will be deposited, see below, had the same dimensions.) The pair-distribution function confirmed amorphization. The amorphous sample was then relaxed at 0 K and subsequently put on top of the crystalline sample. The amorphous and crystalline samples were brought together at an initial distance of about 2 Å, and minimized, while the bottom 3 layers of the crystalline sample, i.e. those layers further apart from the interface, were kept immobile. Subsequently, except for the bottom three layers, the whole system was heated up to 300K and let to equilibrate for 500ps at this temperature using the Nosé-Hoover thermostat. At this point, a block of 20 Å width, 10 Å height, and 108 Å length of the amorphous sample, which was placed right at the interface, was heated up to temperatures high enough for crystallization to take place during a period of 1ns with a timestep of 1fs; the rest of the system was kept at 300 K (except for the bottom 3 layers of the crystalline sample, which remained immobile). Raising the temperature of the block is taken here to mimic the effect that a beam focused in the block might produce. Once crystallization was observed, the heated region was moved up by 5 Å and the heating process was repeated until the crystallization front reached approximately half the height of the amorphous sample. All the calculations were performed with the software LAMMPS[43,44] and run in the supercomputer *Titan*.

**Amorphous Si growth:**

Two sets of samples were used in this work. For the first experiment, the amorphous silicon layer was deposited on a single crystal silicon wafer shortly after its surface was RCA cleaned. The amorphous silicon deposition process is described elsewhere.[45] For the second, the Si substrates were prepared in ultrahigh vacuum with a base pressure of $4.5 \times 10^{-11}$ Torr. To prepare a Si(100)-(2x1) reconstruction and atomically flat morphology, the substrates were degassed and flash-annealed according to commonly established procedures. The surface quality was examined by STM and LEED. A 12 nm thick amorphous Si film was deposited onto these Si(100) substrates at room temperature using an e-beam evaporator in ultrahigh vacuum. The sample was exposed to the ambient conditions before STEM sample preparation. The Bi-doped Si heterostructure was grown in multiple steps to reduce the Bi segregation from the Si. Specifically, we first synthesized Bi nanolines according to Ref. [46] by evaporating Bi from an effusion held cell at 485 °C onto a (2×1)-reconstructed Si(100) substrate held at 570 °C. A thin crystalline Si layer was subsequently grown by solid phase epitaxy: a thin amorphous Si film was deposited at room temperature and then annealed at 434 °C for 5 s during which it crystallized. Subsequently, a 24 nm thick amorphous Si was deposited at room temperature.

21	Wang, Z. L., Itoh, N., Matsunami, N. & Zhao, Q. T. ION-INDUCED CRYSTALLIZATION AND AMORPHIZATION AT CRYSTAL/AMORPHOUS INTERFACES OF SILICON. *Nucl. Instrum. Methods Phys. Res. Sect. B-Beam Interact. Mater. Atoms* **100**, 493-501, doi:10.1016/0168-583x(95)00369-x (1995).

22	Jencic, I., Bench, M. W., Robertson, I. M. & Kirk, M. A. ELECTRON-BEAM-INDUCED CRYSTALLIZATION OF ISOLATED AMORPHOUS REGIONS IN SI, GE, GAP, AND GAAS. *Journal of Applied Physics* **78**, 974-982 (1995).

23	Yang, X. X., Wang, R. H., Yan, H. P. & Zhang, Z. Low energy electron-beam-induced recrystallization of continuous GaAs amorphous foils. *Mater. Sci. Eng. B-Solid State Mater. Adv. Technol.* **49**, 5-13, doi:10.1016/s0921-5107(97)00104-9 (1997).

24	Li, Z. C., Liu, L., He, L. L. & Xu, Y. B. Electron-beam induced nucleation and growth in amorphous GaAs. *Acta Metall. Sin.* **39**, 13-16 (2003).

25	Egerton, R. Mechanisms of radiation damage in beam-sensitive specimens, for TEM accelerating voltages between 10 and 300 kV. *Microscopy research and technique* **75**, 1550-1556 (2012).

26	Egerton, R., Li, P. & Malac, M. Radiation damage in the TEM and SEM. *Micron* **35**, 399-409 (2004).

27	Belianinov, A. *et al.* Big data and deep data in scanning and electron microscopies: deriving functionality from multidimensional data sets. *Advanced Structural and Chemical Imaging* **1**, 6 (2015).

28	Vasudevan, R. K. *et al.* Big data in reciprocal space: Sliding fast Fourier transforms for determining periodicity. *Applied Physics Letters* **106**, 091601 (2015).

29	Seitz, F. & Koehler, J. Displacement of atoms during irradiation. *Solid State Physics-Advances In Research And Applications* **2**, 305-448 (1956).

30	Kaganov, M., Lifshitz, I. & Tanatarov, L. Relaxation between electrons and the crystalline lattice. *Sov. Phys. JETP* **4**, 173-178 (1957).

31	Meftah, A. *et al.* Track formation in SiO 2 quartz and the thermal-spike mechanism. *Physical Review B* **49**, 12457 (1994).

32	Wong, B. T. & Mengüç, P. M. *Thermal Transport for Applications in Micro/Nanomachining*. (Springer Science & Business Media, 2008).

33	Chettah, A. *et al.* Behavior of crystalline silicon under huge electronic excitations: A transient thermal spike description. *Nuclear Instruments and Methods in Physics Research Section B: Beam Interactions with Materials and Atoms* **267**, 2719-2724 (2009).

34	Osmani, O. *et al.* Damage in crystalline silicon by swift heavy ion irradiation. *Nuclear Instruments and Methods in Physics Research Section B: Beam Interactions with Materials and Atoms* **282**, 43-47 (2012).

35	Daraszewicz, S. & Duffy, D. Extending the inelastic thermal spike model for semiconductors and insulators. *Nuclear Instruments and Methods in Physics Research Section B: Beam Interactions with Materials and Atoms* **269**, 1646-1649 (2011).

36	Khara, G. S., Murphy, S. T., Daraszewicz, S. L. & Duffy, D. M. The influence of the electronic specific heat on swift heavy ion irradiation simulations of silicon. *Journal of Physics: Condensed Matter* **28**, 395201 (2016).

37	Akkerman, A. & Murat, M. Electron–phonon interactions in silicon: Mean free paths, related distributions and transport characteristics. *Nuclear Instruments and Methods in Physics Research Section B: Beam Interactions with Materials and Atoms* **350**, 49-54 (2015).

38	Bell, R., Toulemonde, M. & Siffert, P. Calculated temperature distribution during laser annealing in silicon and cadmium telluride. *Applied Physics A: Materials Science & Processing* **19**, 313-319 (1979).
19